\documentclass[a4paper]{article}

\usepackage{INTERSPEECH2020}
\usepackage{multirow}
\usepackage{wrapfig}
\usepackage{phonetic}
\usepackage{cite}
\usepackage{pdfpages}
\usepackage{makecell}
\usepackage{tabularx}
\usepackage{tipa}
\usepackage{enumitem}
\usepackage{array}
\newcolumntype{x}[1]{>{\centering\arraybackslash\hspace{0pt}}p{#1}}

\title{Classification of Manifest Huntington Disease using Vowel Distortion Measures}
\name{Amrit Romana$^1$, John Bandon$^1$, Noelle Carlozzi$^2$, Angela Roberts$^3$, Emily Mower Provost$^1$}
\address{
  $^1$Computer Science and Engineering, University of Michigan, Ann Arbor, Michigan, USA\\
  $^2$Physical Medicine \& Rehabilitation, University of Michigan, Ann Arbor, Michigan, USA\\
  $^3$Communication Sciences and Disorders, Northwestern University, Evanston, Illinois, USA}
\email{aromana@umich.edu, jbandon@umich.edu, carlozzi@med.umich.edu, angela.roberts@northwestern.edu, emilykmp@umich.edu}

\begin{document}

\maketitle

\begin{abstract}
Huntington disease (HD) is a fatal autosomal dominant neurocognitive disorder that causes cognitive disturbances, neuropsychiatric symptoms, and impaired motor abilities (e.g., gait, speech, voice). Due to its progressive nature, HD treatment requires ongoing clinical monitoring of symptoms. Individuals with the gene mutation which causes HD may exhibit a range of speech symptoms as they progress from premanifest to manifest HD. Differentiating between premanifest and manifest HD is an important yet understudied problem, as this distinction marks the need for increased treatment. Speech-based passive monitoring has the potential to augment clinical assessments by continuously tracking manifestation symptoms. In this work we present the first demonstration of how changes in connected speech can be measured to differentiate between premanifest and manifest HD. To do so, we focus on a key speech symptom of HD: vowel distortion. We introduce a set of vowel features which we extract from connected speech. We show that our vowel features can differentiate between premanifest and manifest HD with 87\% accuracy. 




\end{abstract}

\section{Introduction}

Huntington disease (HD) is a fatal autosomal dominant neurocognitive disorder that causes cognitive disturbances, neuropsychiatric symptoms, and impaired motor abilities (e.g., gait, speech, voice) \cite{vonsattel1998huntington, snowden2001longitudinal, paulsen2001neuropsychiatric, long2014tracking}.
Individuals who have a family history of HD can undergo a gene test to learn if they carry the gene mutation that causes HD (i.e., are gene-positive). Individuals who are gene-positive will develop clinically significant symptoms of HD, resulting in an HD diagnosis, typically in their mid-40's \cite{duyao1993trinucleotide}. These individuals are considered premanifest before the onset of these symptoms, and manifest after. No cure exists, but timely diagnosis of HD (i.e., manifestation) coupled with treatment allows individuals to manage their symptoms. 

At-home passive symptom monitoring captures patient health as it relates to real-world functioning \cite{kabelac2019passive}.
Providing clinicians with this information can allow for a more timely diagnosis of HD and a better understanding of its progression for treatment planning. 
Disordered speech is one symptom of HD, and previous work has demonstrated changes in speech occur before an HD diagnosis, and become more noticeable as HD progresses. \cite{kaploun2011acoustic, vogel2012speech, hinzen2018systematic, perez2018classification}. This suggests the potential of passively tracking speech symptoms to better understand HD progression.    

Vowel distortion is one speech symptom of HD \cite{darley1969differential, hartelius2003speech, diehl2019motor}, and tracking this symptom may augment passive monitoring. However, methods of automatically quantifying vowel distortion from connected speech have not been extensively explored. Kaploun et al. extracted jitter and shimmer from a sustained vowel task to characterize vowel distortion, and they demonstrated its prevalence as an HD symptom \cite{kaploun2011acoustic}. Works differentiating between healthy and disordered speech (a range of conditions in the Kay Elemetrics Disordered Voice Dataset \cite{elemetrics1994disordered}) have extracted measures of system stability from sustained vowel tasks, suggesting the potential of stability measures for capturing vowel distortion  \cite{little2007exploiting, henriquez2009characterization, vaziri2010pathological, elemetrics1994disordered}. However, measures extracted from sustained vowel tasks may not have the same information about speech or voice disorders when extracted from vowels in connected speech. Vowels in connected speech differ 
because they are 1) modified and often nonstationary (i.e., changing mean, variance, and frequency properties) due to coarticulation and 2) shorter, which may pose problems for distortion measures that rely on lengthy signals. Thus, to incorporate tracking of vowel distortion into passive speech monitoring, we must assess how these measures relate to HD when extracted from connected speech. In this work we analyze read speech, which is one type of connected speech and includes short vowel samples that are modified due to coarticulation.

The novelty of this work is a new set of features which account for the characteristics of connected speech and reliably measure vowel distortion as it relates to HD. We present the first system to classify premanifest versus manifest HD using features from connected speech, doing so with 87\% accuracy.


\section{Related work}\label{sec:related_work_vowel}

\subsection{HD classification using speech}
Previous works have demonstrated the potential of passively monitoring speech to assist in managing neurocognitive disorders such as Parkinson's
\cite{ tsanas2012novel, orozco2016automatic} and Alzheimer's
\cite{ satt2014speech, konig2015automatic, zhou2016speech}. Individuals with HD exhibit similar speech symptoms, suggesting the potential of monitoring speech to aid in managing HD.

The majority of work in classifying HD stages using speech has not studied how to differentiate between premanifest and manifest HD, but they have differentiated between healthy controls and individuals who are gene-positive \cite{kaploun2011acoustic, perez2018classification}. Kaploun et al. used speaking rate from a reading passage and jitter, shimmer, and the noise to harmonics ratio of a sustained vowel, to classify individuals as healthy controls or premanifest \cite{kaploun2011acoustic}. 
This work illustrates that subtle speech symptoms occur even in the premanifest population. Perez et al. used speaking rate, filler frequency, pause information, and goodness of pronunciation features from a reading passage to classify individuals as healthy controls or gene-positive with 87\% accuracy \cite{perez2018classification}. In doing so, Perez et al. demonstrated the difficulty of differentiating between the premanifest and manifest subcategories: half of premanifest individuals were classified as healthy controls, and the other half as gene-positive. 

\subsection{Vowel stability} 

Prior works have applied additional measures from nonlinear dynamical systems to quantify vowel stability from sustained vowels. These measures have been used to classify a range of disorders and may be applicable in classifying manifest HD. 

Vaziri et al. extracted the correlation dimension (CD) and the Maximal Lyapunov Exponent (MLE) from sustained vowels, and used these measures to classify voice disorders in the Kay Dataset \cite{vaziri2010pathological}. Little et al. highlight nonstationarity as an issue when extracting the CD and MLE from speech. To address this, Little et al. use the detrended fluctuation analysis (DFA) exponent. The DFA exponent characterizes the roughness of noise or fluctuations around vocal cord vibrations and is intended for use with nonstationary data. They extract DFA exponent from sustained vowels to classify voice disorders in the Kay Dataset \cite{little2007exploiting}. In later work, Little et al. also demonstrated how the DFA exponent could be used to measure the severity of Parkinson's disorder \cite{little2008suitability}.

Bryce et al. describe how DFA, although intended to work with nonstationary data, can still fail to detrend in many cases \cite{bryce2012revisiting}. When a series has strong underlying periodicity, DFA introduces artifacts that distort the DFA exponent. Bryce et al. suggest explicitly detrending the signal before analyzing fluctuations. In this paper, we analyze whether DFA effectively detrends vowels, and explore explicitly remove trends before analyzing fluctuations. 




\section{Data description}
We use data collected as part of a study on acoustic biomarkers for HD at the University of Michigan. The study participants provided speech samples that were recorded at 44.1 kHz with a Hosa XVM-102M XLR microphone. We use two tasks: the 
sustained vowel, in which participants were instructed to hold the vowel /a/ for as long as possible, 
and the Grandfather Passage (GFP). The GFP contains nearly all of the phonemes of American English and is a standard reading passage used in assessing motor speech and voice disorders \cite{darley1975motor}.

The data contains speech from 62 individuals, in which 31 are healthy controls and 31 are gene-positive. Gene-positive individuals are assigned to specific HD stages (premanifest, manifest early-stage, and manifest late-stage) using the Unified Huntington's Disease Rating Scale (UHDRS) \cite{kieburtz2001unified}. First, the premanifest versus manifest labels are determined based on the clinician-determined Diagnostic Confidence Level (DCL) within the Total Motor Score (TMS) portion of UHDRS. DCL ranges from 0 (no symptoms) to 4 (symptoms of HD with $>$99\% confidence). We label participants with a DCL of less than 4 as premanifest, and participants with a DCL of 4 as manifest, as in \cite{liu2015motor}. Within the manifest group, we label participants as early- or late-stage based on their Total Functional Capacity (TFC) scores \cite{shoulson1989assessment}. TFC scores provide a rating of functional capacity, and range from 0 (low functioning) to 13 (high functioning). We label participants with a TFC score of 7-13 early-stage and those with a TFC score of 0-6 as late-stage, as in \cite{marder2000rate}.

This paper focuses on analyzing speech of gene-positive individuals. Of these participants, one was unable to hold the sustained vowel. To provide a consistent comparison across experiments, we exclude this participant from our analysis. 
Thus, in this paper, we use data collected from 30 individuals: 12 premanifest, 11 early-stage HD, and 7 late-stage HD. We focus on the binary problem of differentiating between individuals with premanifest HD and manifest HD. 


\section{Methods}

\subsection{Data segmentation}

We segment three vowel sample types: sustained vowels, shortened sustained vowels, and vowels extracted from the GFP. Table \ref{tab:sample_descriptions} summarizes these samples.

We analyze the sustained vowel recordings in which participants were instructed to hold the vowel /a/ for as long as possible after the interviewer provided an example. Recordings varied in length (12.8s $\pm$ 8.5s), as some participants held the vowel for longer than others. For each participant, we segment their sustained vowel (\textbf{SV}) sample.    

To analyze vowel changes within read speech, we manually segment the vowels from GFP recordings. We choose to focus on a single phone to limit potential variation due to sound. We focus specifically on the phone [\openo], as it closely resembled the sounds in the SV samples and, according to the Carnegie Mellon University Pronouncing Dictionary \cite{cmudict} phonetic translation of the GFP, [\openo] has 12 occurrences within the passage, such as the /a/ in the words ``all'' and ``walk''. We listen to each GFP recording for the occurrences of [\openo] identified in the phonetic transcript, and then determine phone endpoints by assessing changes in the sound and associated spectrogram. We verify that the sound of the resulting sample minimally contained surrounding phones. The number and length of samples vary slightly by each participant, as there were variations in speaking rate and pronunciation, making some occurrences of the phones difficult to segment. Ultimately, we extract 11.3 $\pm$ 0.89 vowel samples for each participant, and these are 105ms $\pm$ 49ms in length. We refer to these samples as GFP vowels (\textbf{GFPV}). 

To understand the impact of vowel length versus vowel changes within read speech, we sample an intermediate set of vowels from SV, but of a length representative of GFPV. We sample 10 shortened sustained vowels (\textbf{SSV}) from each SV. The start positions and lengths of the SSV are randomly chosen. The lengths are chosen from a normal distribution with a mean of 105 ms and standard deviation of 49 ms, to resemble the lengths of the GFPV.

\begin{table}[h]
  \caption{Three types of vowel samples for each individual}
  \label{tab:sample_descriptions}
  \centering
  \begin{tabular}{ c c }
    \toprule
    \textbf{Sample type} & \textbf{Description} \\
    \midrule
    SV & 
    \begin{tabular}{@{}c@{}} 
    1 sample holding the vowel /a/ \end{tabular}    \\
    \midrule
    SSV
      & \begin{tabular}{@{}c@{}} 
    10 randomly selected segments from\\
    the SV, each 105 $\pm$ 49 ms in length
    \end{tabular}                    \\
    \midrule
    GFPV & 
    \begin{tabular}{@{}c@{}} 
    11.3 $\pm $0.89 occurrences of the phone {[}\openo{]} \\ 
    manually segmented from the \\ 
    Grandfather Passage reading, \\
    each 105 $\pm$ 49 ms in length
    \end{tabular}                    \\
    
    \bottomrule
  \end{tabular}
\end{table}

\subsection{Feature extraction}

\textbf{Baseline features.} \label{sec:baseline_features}
We develop a set of baseline features for the task of classifying premanifest versus manifest HD. In prior work, Perez et al. extracted 358 features relating to speaking rate, pauses, goodness of pronunciation, and filler usage from the GFP \cite{perez2018classification}. These features were extracted by force aligning audio with manually-created transcripts. Using these features, Perez et al. differentiated between healthy controls and gene-positive individuals with 87\% accuracy, but did not focus on separating the premanifest and manifest populations. 


\textbf{Vowel features overview. }
In the remainder of this section, we describe the extraction of vowel-specific features we explore: vowel length, the original implementation of a DFA exponent, and our proposed trend and fluctuation features. In our preliminary analysis we also compared the use of jitter, shimmer, and the noise to harmonics ratio from vowels within connected speech, but found that these features were not correlated with HD manifestation. Thus in this paper we focus on vowel length, the original implementation of a DFA exponent, and  our proposed features. 

We extract these features from SV, SSV, and GFPV. When extracting features from SV, we have one value per individual. When extracting features from SSV and GFPV, we have one value per vowel. For each individual we aggregate the features of all vowels with six statistics: minimum, median, maximum, range, mean, and standard deviation. 

\textbf{Vowel length.} 
We extract the length of vowels in milliseconds for SV and GFPV. We do not do so for SSV, because the lengths of these samples were randomly determined. We hypothesize that within SV, individuals with manifest HD have shorter vowel lengths. We hypothesize that with GFPV, individuals with manifest HD have longer vowel lengths due to the HD symptom of prolonged sounds \cite{diehl2019motor}.

\textbf{Detrended fluctuation analysis.} 
Detrended fluctuation analysis (DFA) is a method for analyzing the stability of fluctuations. A DFA exponent describes how deviations from a trend increase as we change the scale (or the size of the window) at which we view them. We follow \cite{little2007exploiting} to extract the DFA exponent from our vowel samples, and describe the potential pitfalls of DFA. Little et al. extracted DFA exponents from sustained phonation recorded at 25 kHz \cite{little2007exploiting}. They used linear detrending within each window, and these windows ranged in size from 50 samples (2 ms) to 100 samples (4 ms). 
In extracting DFA exponents, rather than downsampling our data, which may obscure some of the fluctuations we are aiming to characterize, we scale the window sizes so that our windows capture the same temporal information as captured in \cite{little2007exploiting}. Our smallest windows are 88 samples and our largest windows are 176 samples. 

\begin{figure}[h]
    \centering
    \includegraphics[scale=0.6]{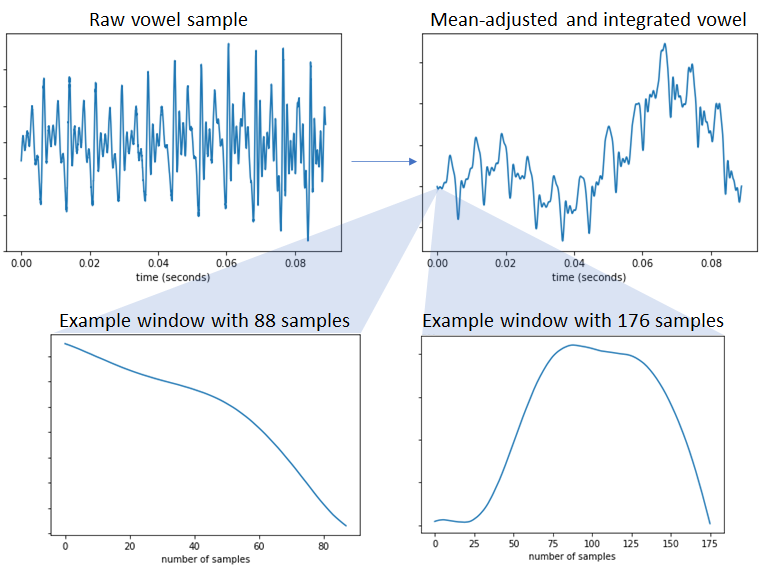}
    \caption{An example of a vowel and the first step of DFA which includes mean-adjusting and integrating the vowel. We highlight the signal within windows of 88 and 176 samples of the integrated vowel. While the smallest window (88 samples) has a clear linear trend, the larger window (176 samples) does not. This suggests that linear detrending may not be effective.}
    \label{fig:window_probs}
\end{figure}

Bryce et al. describe how DFA, although intended to work with nonstationary data, can still fail to detrend in many cases \cite{bryce2012revisiting}. In particular, when a series has strong periodic component, as speech does, DFA may not be able to effectively detrend within each window. Attempting to detrend using DFA may introduce artifacts that distort the DFA exponent. We explore this possibility by looking at speech within various windows, and find that, especially as the windows increase in length, the underlying trends are not linear. This is illustrated with an example in Figure \ref{fig:window_probs}.  To prevent DFA from  introducing artifacts as it tries to detrend, Bryce et al. suggest 
explicitly detrending data before performing fluctuation analysis. This fluctuation analysis will return an estimate of the Hurst exponent (HE), which corresponds with the DFA exponent although the HE is intended for detrended series. We explore this option next. 

\begin{figure*}[t]
\centering
    \includegraphics[scale=0.5]{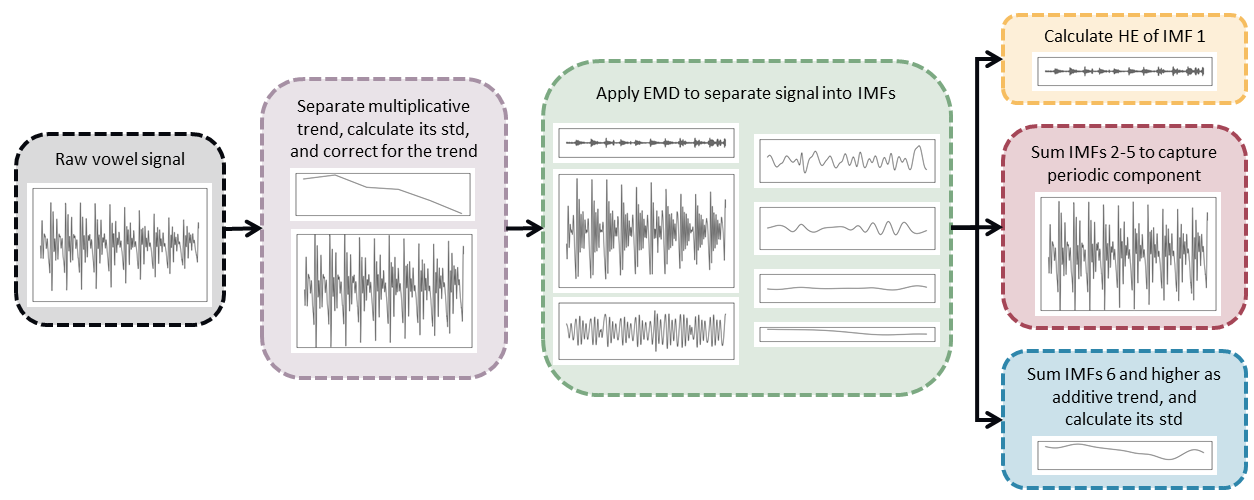}
\vspace*{-2mm}
\caption{Pipeline for extracting three vowel measures: 1) standard deviation of the multiplicative trend; 2) standard deviation of the additive trend; and 3) HE of the first IMF. EMD=empirical model decomposition, IMF=intrinsic mode function, HE=Hurst exponent}
\label{fig:VDM}
\end{figure*}

\textbf{Trend and fluctuation measures. }
We propose a pipeline for explicitly detrending a vowel before performing fluctuation analysis. This method is reliant on empirical mode decomposition (EMD), which is an adaptive sifting method that works in the time domain and separates a signal into intrinsic mode functions (IMFs) that have varying frequency content. The first IMF contains the highest frequency content, and thus contains the noise-like fluctuations in the signal. As we remove the trends, we also quantify them to explore their relevance to HD manifestation. Previous work has suggested that nonstationarities introduced in connected speech result from movement of articulators, such as the tongue and lips \cite{little2007exploiting}. Because HD limits these movements, we suspect individuals with manifest HD may exhibit different nonstationarities than individuals with premanifest HD. This approach, summarized in Figure \ref{fig:VDM}, provides three measures for each vowel: the standard deviation of the multiplicative trend, the standard deviation of the additive trend, and an estimate of the HE of the noise-like fluctuations \footnote{Code available at https://github.com/amritkromana/FVDM}. The remainder of this section will describe these steps.  

\textit{Quantifying and removing the multiplicative trend.} A multiplicative trend in speech indicates changes in volume. Previous work has highlighted monoloudness as a symptom of HD \cite{diehl2019motor}. However, individuals with manifest HD also exhibit prolonged sounds, which may result in an increase in volume changes around vowels. We address this trend by first calculating the average decibels relative to full scale (dBFS, a measure of amplitude) of the vowel. We then apply a filter, with a window of 25 ms and a shift of 10 ms, to calculate the average dBFS within each window. We calculate the standard deviation of these dBFS values and save it as a feature: the \textbf{standard deviation of the multiplicative trend}. Finally, we correct for this trend by applying the necessary gain or decay to each window so that it matches the average dBFS of the entire vowel. The goal of removing this trend is to work toward a detrended signal from which we can analyze fluctuations. 

\begin{figure}[tp]
\centering
    \includegraphics[scale=0.5]{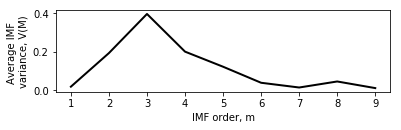}
\vspace*{-2mm}
\caption{IMF Variance for random vowel samples}
\label{fig:IMF_var}
\end{figure}

\textit{Separating the signal into IMFs.} In order to avoid conflating volume information in the rest of our analysis, we normalize the amplitude of each signal to [-1,1]. We then apply EMD to separate the signal into IMFs \cite{emd2017}. The first IMF will have the highest frequency content, and later IMFs will have lower frequency content. The main advantage of this filtering approach is that it does not make assumptions about the type of periodicity. 

\textit{Quantifying the additive trend.} An additive trend in speech is a potential artifact of coarticulation due to movement of the articulators. We hypothesize that individuals with manifest HD exercise less articulator movement, which may be evident in fewer changes in this trend.  Chatlani et al. explore IMF characteristics of voiced sounds, and they provide methods to associate certain IMFs with signal information and certain IMFs with low-frequency trend information \cite{chatlani2011emd}. They demonstrated how the variance of each IMF drops after the fourth IMF, and suggest that the first four IMFs contain relevant signal information. We repeat this analysis using 50 random vowel samples from our dataset. Figure \ref{fig:IMF_var} illustrates that within our data the first five IMFs have higher variance, after which variance drops. This difference in which IMFs contain signal (the first four versus the first five) may be due to different recording conditions or the fact that our dataset includes disordered speech. 
Based on this analysis, we sum IMFs higher than five as the additive trend. We calculate the standard deviation of this trend and save it as a feature: the \textbf{standard deviation of the additive trend}.

\textit{Analyzing fluctuations.} The first IMF contains the highest frequency content, while the higher IMFs are more likely to contain periodic components. Thus we focus on capturing fluctuations within the first IMF. 
The third vowel feature we introduce is the \textbf{Hurst exponent (HE) of the first IMF}. A HE closer to 1 indicates more smoothness, whereas a HE closer to 0 indicates more roughness. We hypothesize that the HE of the first IMF will be lower for individuals with manifest HD compared to individuals with premanifest HD due to vowel distortion. 

To improve the robustness of our measure, we make the following three modifications to our fluctuation analysis compared to that in \cite{little2007exploiting}. First, we look for deviations from the mean within each window, rather than applying any linear detrending. Next, we expand the maximum window size as recommended in \cite{bryce2012revisiting}. We make the assumption that individuals have a fundamental frequency of at least 100 Hz, and we set the largest window to 441 samples (10 ms) to capture fluctuations across each individual's vocal cord vibration cycle. Finally, we include a filtering step to ensure we are reliably estimating the HE. Measuring the HE involves calculating deviations within windows of various sizes, creating a log-log plot of these deviations versus the size of the window, and then taking the slope of a best-fit line to the data. In general, one can always fit a line to the log-log plot, but the slope of this line is only an estimate of the HE if the data are linear and the line is a good fit. In particular, for our SSV and GFVP samples where we have multiple samples for each individual, we drop vowel samples if the line fit to the log-log plot has an $R^2$ of less than 0.99. Table \ref{tab:r2_counts} displays the number of vowel samples used in the HE estimates for each individual before and after this reduction, and illustrates that the majority of vowels satisfy these requirements. However, this reduction in samples may limit the usefulness of certain statistics, such as the range and standard deviation. In future work we will continue to evaluate what vowel characteristics lead to certain vowels not producing linear behavior within log-log plots.

\begin{table}
  \caption{Number of vowel samples per individual before and after reduction for $R^2$ $>$ 0.99. Counts displayed by sample type and manifest label. Note: SSV=shortened sustained vowels, GFPV=Grandfather Passage vowels.}
  \label{tab:r2_counts}
  \centering
  \begin{tabular}{c c c c}
    \toprule
    & & \textbf{Premanifest HD} & \textbf{Manifest HD} \\
    \midrule
    \multirow{2}{*}{SSV} & Before & 10 $\pm$ 0 & 10 $\pm$ 0\\
    & After & 8.42 $\pm$ 1.44 & 6.67 $\pm$ 2.59\\ 
    \multirow{2}{*}{GFPV} & Before & 11.42 $\pm$ 0.97 & 11.28 $\pm$ 0.83\\ 
    & After & 7.33 $\pm$ 2.53 & 6.22 $\pm$ 2.62\\ 
    \\[-1em]
  \bottomrule
  \end{tabular}
\end{table}

In summary, we propose three features: 
\begin{enumerate}
    \item Standard deviation of the multiplicative trend, which we expect will be higher for individuals with manifest HD due to prolonged sounds
    \item Standard deviation of the additive trend, which we expect may be lower for individuals with manifest HD due to less articulator movement
    \item HE of the first IMF, which we expect will be lower for individuals with manifest HD due to distortion in the vowel 
\end{enumerate}
\begin{table*}[t]
  \caption{Spearman correlation coefficients between vowel features and manifest label for each sample type. Significant correlations (p $<$ 0.05) are in bold.  Note:  SV=sustained vowel, SSV=shortened sustained vowels, GFPV=Grandfather Passage vowels, DFA=Detrended fluctuation analysis, HE=Hurst exponent, IMF=intrinsic mode function, SD=standard deviation.}

  \label{tab:feature_correlations}
  \centering

\begin{tabular}{ x{1.1cm} x{0.75cm} x{1.45cm} x{1.45cm} x{1.45cm} x{1.45cm} x{1.45cm}}
    \toprule
    \multirow{3}{*}{\textbf{Sample}} &
    \multirow{3}{*}{\textbf{Stat}} &
    \multirow{3}{*}{\thead{\textbf{Vowel} \\ \textbf{length}}} &
    \multirow{3}{*}{\thead{\textbf{DFA} \\ \textbf{exponent}}} &
    \multicolumn{3}{c}{\textbf{Trend and fluctuation measures}} \\
    & & & & \thead{\textbf{HE of } \\ \textbf{first IMF}} & \thead{\textbf{SD of the} \\ \textbf{ mult. trend}} & \thead{\textbf{SD of the} \\ \textbf{add. trend}} \\ 
    \midrule 
    SV & & \textbf{-0.48} & -0.23 & -0.08 & \textbf{0.42} & \textbf{-0.51} \\ 
    \midrule 
    \multirow{6}{*}{SSV} & 
    min & - & -0.30 & -0.24 & 0.08 & -0.31 \\
    & med & - & -0.27 & -0.26 & 0.21 & \textbf{-0.45} \\
    & max & - & -0.04 & -0.13 & 0.20 & -0.19 \\
    & range & - & 0.31 & 0.11 & 0.20 & -0.12 \\
    & mean & - & -0.28 & -0.20 & 0.21 & \textbf{-0.36} \\ 
    & SD & - & 0.35 & 0.06 & 0.20 & -0.13 \\
    \midrule
    \multirow{6}{*}{GFPV} & min & 0.09 & -0.17 & \textbf{-0.39} & 0.06 & \textbf{-0.36} \\
    & med & \textbf{0.39} & -0.20 & \textbf{-0.53} & 0.17&-0.31 \\
    & max & \textbf{0.52} & -0.06 & \textbf{-0.39} & \textbf{0.55} & -0.16 \\ 
    & range & \textbf{0.56} & 0.21 & -0.12 & \textbf{0.55} & -0.13 \\ 
    & mean & \textbf{0.42} & -0.17 & \textbf{-0.53} & \textbf{0.38} & -0.28 \\ 
    & SD & \textbf{0.47} & 0.14 & 0.02 & \textbf{0.50} & -0.11    \\ 
        \bottomrule
  \end{tabular}
\end{table*}

\section{Results}

\subsection{Feature correlations across sample types}

Table \ref{tab:feature_correlations} lists the correlations between the vowel measures extracted from each sample type and the binary manifest labels. Note that we do not extract vowel length from SSV, as those lengths were randomly chosen. We highlight the correlations greater than 0.5 between several vowel features (specifically vowel length, the HE of the first IMF, and the standard deviation of the multiplicative trend) and the manifest label when these features are extracted from vowels in connected speech (GFPV). We further analyze these findings in this section. 

Vowel length is negatively correlated with HD manifestation when extracted from SV but positively correlated with HD manifestation when extracted from GFPV. This is consistent with HD symptoms, including prolonged sounds \cite{diehl2019motor}. 

The DFA exponent, extracted using the method outlined in \cite{little2007exploiting}, is not significantly correlated with HD manifestation when extracted from any of the vowel samples. This suggests that this measure may not be measuring fluctuations accurately in our dataset, potentially due to detrending problems.

The HE of the first IMF is significantly correlated with HD manifestation when extracted from GFPV, but not SSV or SV. For longer samples, we find that EMD separates the signal into a much larger number of IMFs. For SV, the noise is potentially contained within higher IMFs as opposed to just the first IMF, and future work will explore this possibility. For SSV, we do not find significant correlations, but we find that the direction of these correlations is consistent with GFPV. This suggests that vowel distortion is more pronounced within read speech. Within the GFPV we find that the median and mean of this feature are correlated with HD manifestation with a coefficient of -0.53.

The standard deviation of the multiplicative trend is positively correlated with HD manifestation when measured from all the samples, and significantly so when extracted from SV and GFPV. 
We suspect the higher correlations with GFPV samples may be related to the HD symptoms of prolonged sounds and an increase in pauses, which imply phones are more likely to exhibit volume variations.  

Finally, we find significant correlations between the standard deviation of the additive trend and HD manifestation. The negative correlation between the minimum of this feature from GFPV and manifest HD is consistent with less articulator movement due to disease manifestation. However, we find this negative correlation within SSV where we do not expect articulator movement. Future work will aim to analyze the factors driving these correlations. Finally, we note the high correlation between this measure when extracted from SV. Again, we believe this is due to differences in IMF characteristics across shorter and longer samples. The SV samples were decomposed into a higher number of IMFs, so IMF five and higher likely contain different frequencies in SV compared to SSV and GFPV.

\subsection{Classifying manifest HD}

We explore the feasibility of detecting HD manifestation from speech by training a logistic regression model to classify a speaker as having premanifest or manifest HD. We train the model using a leave-one-subject-out paradigm, meaning that for each participant, we train a model using data from all of the other participants, and use this model to classify the held out participant.  We implement the logistic regression model using scikit-learn \cite{scikit-learn}. We use the Limited-memory Broyden–Fletcher–Goldfarb–Shannon (LBFGS) solver with L2 regularization (C=1.0). We perform z-score normalization on each of the features, using the mean and standard deviation of the training set. We also perform feature selection using the training set. We limit the model to using ten features, and choose those features to maximize relevance to the label while minimizing multicollinearity. More specifically, we begin with a list of potential features which are correlated with the manifest label, indicated by a Spearman correlation p-value of less than 0.1. We then include the feature with the strongest correlation coefficient in our confirmed feature set. We then calculate the variance inflation factor (VIF) between the confirmed feature set and each feature in the potential feature list. We remove features from the potential feature list if their VIF is greater than 5, as this indicates multicollinearity \cite{kim2019multicollinearity}. With our reduced potential feature list, we move the one feature with the strongest correlation coefficient to our confirmed feature set. We repeat this process until the confirmed feature set includes ten features or the list of potential features is empty.   

We compare the classification accuracy within and across different feature sets: baseline, vowel, and both combined. We analyze the relevance of baseline features which have been demonstrated to be relevant to separating healthy controls from gene-positive individuals \cite{perez2018classification}. We then assess the accuracy of a model using the vowel features extracted from the GFP. We use vowel features extracted from the GFP because the baseline features were extracted from the GFP, and this provides the most insight into how to passively predict manifestation from connected speech. We do not present results  here for using the DFA exponent, as these features were not significantly correlated with the manifest label and as a result these features were not selected by the feature selector. Finally, we evaluate the accuracy of a model that combines the baseline features with the vowel features. The results for each experiment are in Table \ref{tab:classification_results}.

 \renewcommand\theadfont{}

\begin{table}[t]
  \caption{Accuracy and F1 scores for classifying premanifest vs manifest HD using the baseline features, vowel features, and combining the feature sets. We use our feature selection method to choose 10 features for each experiment, and for the experiment that combines baseline features with vowel features we enforce 5 features from the baseline set and 5 from the vowel set. Best scores for each feature set tested in bold.}
  \label{tab:classification_results}
  \centering
  \begin{tabular}{c c c}
    \toprule
    \textbf{Features} & \textbf{Accuracy} & \textbf{F1 score} \\
    \midrule
    Pauses & 0.67 & 0.71 \\ 
    Rate & 0.63 & 0.67 \\
    GOP & 0.63 & 0.65 \\ 
    Fillers & 0.63 & 0.62 \\ 
    \textbf{All Baseline}  & \textbf{0.73} & \textbf{0.76} \\ 
    \midrule
    Vowel length & 0.73 & 0.76 \\
    Trend + Fluctuation & 0.67 & 0.71 \\ 
    \textbf{All Vowel Features} & \textbf{0.87} & \textbf{0.88} \\
    \midrule
    \thead{\textbf{Baseline Features +} \\ \textbf{Vowel Features}} & \textbf{0.80}	& \textbf{0.83} \\
\bottomrule
  \end{tabular}
\end{table}

The speech rate, goodness of pronunciation, and filler feature sets each perform comparably with 63\% accuracy, and the pause features perform slightly better with 67\% accuracy. We find combining all of the baseline feature sets improves performance, reaching 73\% accuracy. Evaluating the vowel features individually, we find that vowel length performs the best, also with 73\% accuracy. Adding trend and fluctuation features increases this accuracy to 87\%. In combining baseline features with vowel features, we limited our feature selection process to choose up to five features from the baseline set and five features from the vowel set. However, we find that this does not perform as well as the vowel features on their own. Overall, our results suggest that the baseline features, which separated healthy controls from gene-positive individuals with high accuracy \cite{perez2018classification}, are not as relevant to classifying HD manifestation.

We further analyze the features selected and their beta coefficients to understand the relative importance of features and ensure that the model is interpretable. We perform this analysis with our best-performing experiment, which included all vowel features. Table \ref{tab:features_selected} lists the features that were selected in the majority of training folds, as well as the mean and standard deviation of derived coefficients. The range of vowel length has the highest coefficient of 1.11 $\pm$ 0.07. The mean of vowel length also has a positive coefficient, consistent with the HD symptom of having prolonged sounds \cite{diehl2019motor}. The maximum, median, and minimum of the HE of the first IMF all have negative coefficients, consistent with the HD symptom of distorted vowels \cite{diehl2019motor, darley1969differential}. The coefficients for the maximum of the standard deviation of the multiplicative trend is positive, while coefficient of the mean of the standard deviation of this trend is slightly negative. We suspect this may be because one symptom of HD is monoloudness \cite{diehl2019motor}. Finally, we find a negative coefficient for the standard deviation of the additive trends, consistent with our hypothesis of less articulator movement. 

\begin{table}
  \caption{Most common features selected from manifest HD classification experiment with all vowel features. Each of these features was selected in the majority of training folds with significant beta coefficients (p-values $<$ 0.05). We report the mean and standard deviation of the beta coefficients associated with each feature across all training folds.}
  \label{tab:features_selected}
  \centering
  \begin{tabular}{c c}
    \toprule
     \textbf{Feature} & \textbf{$\beta$} \\
    \midrule
    Mean of vowel length & 0.79 $\pm$ 0.07 \\
    Range of vowel length & 1.11 $\pm$ 0.07 \\
    Maximum of HE of first IMF	& -0.56 $\pm$ 0.07 \\ 
    Median of HE of first IMF	& -1.00 $\pm$ 0.06 \\ 
    Minimum of HE of first IMF & -0.28 $\pm$ 0.06 \\ 
    Maximum of SD of multiplicative trend	& 0.77 $\pm$ 0.06 \\ 
    Mean of SD of multiplicative trend & -0.18 $\pm$ 0.08 \\ 
    Minimum of SD of additive trend & -1.00 $\pm$ 0.07 \\
\bottomrule
  \end{tabular}
\end{table}



Overall we find our vowel measures provide information that is supplementary to existing speech features, and improve manifest HD classification accuracy from 73\% to 87\%. Furthermore, we observe that the relationship between the proposed vowel features and HD manifestation is consistent with understandings of vowel changes within HD \cite{darley1969differential, hartelius2003speech, diehl2019motor}.

\section{Conclusions and Future Work}

In this paper, we present a small and interpretable feature set to capture changes in vowels with HD manifestation. We show that these features can classify HD manifestation with 87\% accuracy.  These results bring us closer to being able to passively detect HD manifestation. These features also provide an avenue for understanding the changes in vowels within connected speech as they relate to other neurocognitive disorders. 

In future work, we will focus on coupling these techniques with vowel detection, so we can automatically extract these features. After automating the extraction of these measure, we plan on evaluate their relevance to additional datasets, including understanding vowel characteristics for healthy controls, classifying manifest HD from spontaneous speech, and understanding other neurocognitive conditions. 



Our future work will also explore additional vowel features. Riad et al. recently analyzed characteristics of sustained vowels for predicting HD severity \cite{riad2020vocal}. Several of our findings are consistent with theirs, namely that vowel length is a significant indicator of disease manifestation while DFA is not. Riad et al. also find a number of voice break and MFCC-related features to be related to disease manifestation and severity. In our future work we will explore if these features are similarly related to HD severity when extracted from vowels in connected speech.

\section{Acknowledgements}
We thank the investigators and coordinators of this study, the study participants, the Huntington Study Group, and the Huntington’s Disease Society of America. This work was supported by the National Institutes of Health (NIH), National Center for Advancing Translational Sciences (UL1TR000433), the Heinz C Prechter Bipolar Research Fund, and the Richard Tam Foundation at the University of Michigan. A portion of this study sample was collected in conjunction with NIH, National Institute of Neurological Disorders and Stroke (R01BS077946) and/or Enroll-HD (funded by the CHDI Foundation). 


\bibliographystyle{IEEEtran}

\bibliography{mybib}

\begin{thebibliography}{10}
\providecommand{\url}[1]{#1}
\csname url@samestyle\endcsname
\providecommand{\newblock}{\relax}
\providecommand{\bibinfo}[2]{#2}
\providecommand{\BIBentrySTDinterwordspacing}{\spaceskip=0pt\relax}
\providecommand{\BIBentryALTinterwordstretchfactor}{4}
\providecommand{\BIBentryALTinterwordspacing}{\spaceskip=\fontdimen2\font plus
\BIBentryALTinterwordstretchfactor\fontdimen3\font minus
  \fontdimen4\font\relax}
\providecommand{\BIBforeignlanguage}[2]{{%
\expandafter\ifx\csname l@#1\endcsname\relax
\typeout{** WARNING: IEEEtran.bst: No hyphenation pattern has been}%
\typeout{** loaded for the language `#1'. Using the pattern for}%
\typeout{** the default language instead.}%
\else
\language=\csname l@#1\endcsname
\fi
#2}}
\providecommand{\BIBdecl}{\relax}
\BIBdecl

\bibitem{vonsattel1998huntington}
J.~P.~G. Vonsattel and M.~DiFiglia, ``Huntington disease,'' \emph{Journal of
  neuropathology and experimental neurology}, vol.~57, no.~5, p. 369, 1998.

\bibitem{snowden2001longitudinal}
J.~Snowden, D.~Craufurd, H.~Griffiths, J.~Thompson, and D.~Neary,
  ``Longitudinal evaluation of cognitive disorder in huntington's disease,''
  \emph{Journal of the International Neuropsychological Society}, vol.~7,
  no.~1, pp. 33--44, 2001.

\bibitem{paulsen2001neuropsychiatric}
J.~S. Paulsen, R.~Ready, J.~Hamilton, M.~Mega, and J.~Cummings,
  ``Neuropsychiatric aspects of huntington's disease,'' \emph{Journal of
  Neurology, Neurosurgery \& Psychiatry}, vol.~71, no.~3, pp. 310--314, 2001.

\bibitem{long2014tracking}
J.~D. Long, J.~S. Paulsen, K.~Marder, Y.~Zhang, J.-I. Kim, J.~A. Mills, and
  R.~of~the PREDICT-HD Huntington's Study~Group, ``Tracking motor impairments
  in the progression of huntington's disease,'' \emph{Movement Disorders},
  vol.~29, no.~3, pp. 311--319, 2014.

\bibitem{duyao1993trinucleotide}
M.~Duyao, C.~Ambrose, R.~Myers, A.~Novelletto, F.~Persichetti, M.~Frontali,
  S.~Folstein, C.~Ross, M.~Franz, M.~Abbott \emph{et~al.}, ``Trinucleotide
  repeat length instability and age of onset in huntington's disease,''
  \emph{Nature genetics}, vol.~4, no.~4, pp. 387--392, 1993.

\bibitem{kabelac2019passive}
Z.~Kabelac, C.~G. Tarolli, C.~Snyder, B.~Feldman, A.~Glidden, C.-Y. Hsu,
  R.~Hristov, E.~R. Dorsey, and D.~Katabi, ``Passive monitoring at home: A
  pilot study in parkinson disease,'' \emph{Digital Biomarkers}, vol.~3, no.~1,
  pp. 22--30, 2019.

\bibitem{kaploun2011acoustic}
L.~R. Kaploun, J.~H. Saxman, P.~Wasserman, and K.~Marder, ``Acoustic analysis
  of voice and speech characteristics in presymptomatic gene carriers of
  huntington's disease: biomarkers for preclinical sign onset?'' \emph{Journal
  of Medical Speech-Language Pathology}, vol.~19, no.~2, pp. 49--65, 2011.

\bibitem{vogel2012speech}
A.~P. Vogel, C.~Shirbin, A.~J. Churchyard, and J.~C. Stout, ``Speech acoustic
  markers of early stage and prodromal huntington's disease: a marker of
  disease onset?'' \emph{Neuropsychologia}, vol.~50, no.~14, pp. 3273--3278,
  2012.

\bibitem{hinzen2018systematic}
W.~Hinzen, J.~Rossell{\'o}, C.~Morey, E.~Camara, C.~Garcia-Gorro, R.~Salvador,
  and R.~de~Diego-Balaguer, ``A systematic linguistic profile of spontaneous
  narrative speech in pre-symptomatic and early stage huntington's disease,''
  \emph{Cortex}, vol. 100, pp. 71--83, 2018.

\bibitem{perez2018classification}
M.~Perez, W.~Jin, D.~Le, N.~Carlozzi, P.~Dayalu, A.~Roberts, and E.~M. Provost,
  ``Classification of huntington disease using acoustic and lexical features.''
  in \emph{Interspeech}, 2018, pp. 1898--1902.

\bibitem{darley1969differential}
F.~L. Darley, A.~E. Aronson, and J.~R. Brown, ``Differential diagnostic
  patterns of dysarthria,'' \emph{Journal of speech and hearing research},
  vol.~12, no.~2, pp. 246--269, 1969.

\bibitem{hartelius2003speech}
L.~Hartelius, A.~Carlstedt, M.~Ytterberg, M.~Lillvik, and K.~Laakso, ``Speech
  disorders in mild and moderate huntington disease: Results of dysarthria
  assessments of 19 individuals,'' \emph{Journal of Medical Speech-Language
  Pathology}, vol.~11, no.~1, pp. 1--15, 2003.

\bibitem{diehl2019motor}
S.~K. Diehl, A.~S. Mefferd, Y.-C. Lin, J.~Sellers, K.~E. McDonell,
  M.~de~Riesthal, and D.~O. Claassen, ``Motor speech patterns in huntington
  disease,'' \emph{Neurology}, vol.~93, no.~22, pp. e2042--e2052, 2019.

\bibitem{elemetrics1994disordered}
K.~Elemetrics, ``Disordered voice database,'' 1994.

\bibitem{little2007exploiting}
M.~A. Little, P.~E. McSharry, S.~J. Roberts, D.~A. Costello, and I.~M. Moroz,
  ``Exploiting nonlinear recurrence and fractal scaling properties for voice
  disorder detection,'' \emph{Biomedical engineering online}, vol.~6, no.~1,
  p.~23, 2007.

\bibitem{henriquez2009characterization}
P.~Henr{\'\i}quez, J.~B. Alonso, M.~A. Ferrer, C.~M. Travieso, J.~I.
  Godino-Llorente, and F.~D{\'\i}az-de Mar{\'\i}a, ``Characterization of
  healthy and pathological voice through measures based on nonlinear
  dynamics,'' \emph{IEEE transactions on audio, speech, and language
  processing}, vol.~17, no.~6, pp. 1186--1195, 2009.

\bibitem{vaziri2010pathological}
G.~Vaziri, F.~Almasganj, and R.~Behroozmand, ``Pathological assessment of
  patients’ speech signals using nonlinear dynamical analysis,''
  \emph{Computers in biology and medicine}, vol.~40, no.~1, pp. 54--63, 2010.

\bibitem{tsanas2012novel}
A.~Tsanas, M.~A. Little, P.~E. McSharry, J.~Spielman, and L.~O. Ramig, ``Novel
  speech signal processing algorithms for high-accuracy classification of
  parkinson's disease,'' \emph{IEEE transactions on biomedical engineering},
  vol.~59, no.~5, pp. 1264--1271, 2012.

\bibitem{orozco2016automatic}
J.~Orozco-Arroyave, F.~H{\"o}nig, J.~Arias-Londo{\~n}o, J.~Vargas-Bonilla,
  K.~Daqrouq, S.~Skodda, J.~Rusz, and E.~N{\"o}th, ``Automatic detection of
  parkinson's disease in running speech spoken in three different languages,''
  \emph{The Journal of the Acoustical Society of America}, vol. 139, no.~1, pp.
  481--500, 2016.

\bibitem{satt2014speech}
A.~Satt, R.~Hoory, A.~K{\"o}nig, P.~Aalten, and P.~H. Robert, ``Speech-based
  automatic and robust detection of very early dementia,'' in \emph{Fifteenth
  Annual Conference of the International Speech Communication Association},
  2014.

\bibitem{konig2015automatic}
A.~K{\"o}nig, A.~Satt, A.~Sorin, R.~Hoory, O.~Toledo-Ronen, A.~Derreumaux,
  V.~Manera, F.~Verhey, P.~Aalten, P.~H. Robert \emph{et~al.}, ``Automatic
  speech analysis for the assessment of patients with predementia and
  alzheimer's disease,'' \emph{Alzheimer's \& Dementia: Diagnosis, Assessment
  \& Disease Monitoring}, vol.~1, no.~1, pp. 112--124, 2015.

\bibitem{zhou2016speech}
L.~Zhou, K.~C. Fraser, and F.~Rudzicz, ``Speech recognition in alzheimer's
  disease and in its assessment.'' in \emph{Interspeech}, 2016, pp. 1948--1952.

\bibitem{little2008suitability}
M.~Little, P.~McSharry, E.~Hunter, J.~Spielman, and L.~Ramig, ``Suitability of
  dysphonia measurements for telemonitoring of parkinson’s disease,''
  \emph{Nature Precedings}, pp. 1--1, 2008.

\bibitem{bryce2012revisiting}
R.~Bryce and K.~Sprague, ``Revisiting detrended fluctuation analysis,''
  \emph{Scientific reports}, vol.~2, p. 315, 2012.

\bibitem{darley1975motor}
F.~L. Darley, A.~E. Aronson, and J.~R. Brown, in \emph{Motor speech
  disorders}.\hskip 1em plus 0.5em minus 0.4em\relax W.B. Saunders Company,
  1975.

\bibitem{kieburtz2001unified}
K.~Kieburtz, J.~B. Penney, P.~Corno, N.~Ranen, I.~Shoulson, A.~Feigin,
  D.~Abwender, J.~T. Greenarnyre, D.~Higgins, F.~J. Marshall \emph{et~al.},
  ``Unified huntington’s disease rating scale: reliability and consistency,''
  \emph{Neurology}, vol.~11, no.~2, pp. 136--142, 2001.

\bibitem{liu2015motor}
D.~Liu, J.~D. Long, Y.~Zhang, L.~A. Raymond, K.~Marder, A.~Rosser, E.~A.
  McCusker, J.~A. Mills, J.~S. Paulsen, P.-H. Investigators \emph{et~al.},
  ``Motor onset and diagnosis in huntington disease using the diagnostic
  confidence level,'' \emph{Journal of neurology}, vol. 262, no.~12, pp.
  2691--2698, 2015.

\bibitem{shoulson1989assessment}
I.~Shoulson, R.~Kurlan, A.~J. Rubin, D.~Goldblatt, J.~Behr, C.~Miller,
  J.~Kennedy, K.~A. Bamford, E.~D. Caine, D.~K. Kido \emph{et~al.},
  ``Assessment of functional capacity in neurodegenerative movement disorders:
  Huntington’s disease as a prototype,'' \emph{Quantification of neurologic
  deficit. Boston: Butterworths}, pp. 271--283, 1989.

\bibitem{marder2000rate}
K.~Marder, H.~Zhao, R.~Myers, M.~Cudkowicz, E.~Kayson, K.~Kieburtz, C.~Orme,
  J.~Paulsen, J.~Penney, E.~Siemers \emph{et~al.}, ``Rate of functional decline
  in huntington’s disease,'' \emph{Neurology}, vol.~54, no.~2, pp. 452--452,
  2000.

\bibitem{cmudict}
``\BIBforeignlanguage{english}{Carnegie mellon university pronouncing
  dictionary},'' Carnegie Mellon University.

\bibitem{emd2017}
D.~Laszuk, ``Python implementation of empirical mode decomposition algorithm,''
  \url{http://www.laszukdawid.com/codes}, 2017-.

\bibitem{chatlani2011emd}
N.~Chatlani and J.~J. Soraghan, ``Emd-based filtering (emdf) of low-frequency
  noise for speech enhancement,'' \emph{IEEE Transactions on Audio, Speech, and
  Language Processing}, vol.~20, no.~4, pp. 1158--1166, 2011.

\bibitem{scikit-learn}
F.~Pedregosa, G.~Varoquaux, A.~Gramfort, V.~Michel, B.~Thirion, O.~Grisel,
  M.~Blondel, P.~Prettenhofer, R.~Weiss, V.~Dubourg, J.~Vanderplas, A.~Passos,
  D.~Cournapeau, M.~Brucher, M.~Perrot, and E.~Duchesnay, ``Scikit-learn:
  Machine learning in {P}ython,'' \emph{Journal of Machine Learning Research},
  vol.~12, pp. 2825--2830, 2011.

\bibitem{kim2019multicollinearity}
J.~H. Kim, ``Multicollinearity and misleading statistical results,''
  \emph{Korean journal of anesthesiology}, vol.~72, no.~6, p. 558, 2019.

\bibitem{riad2020vocal}
R.~Riad, H.~Titeux, L.~Lemoine, J.~Montillot, J.~H. Bagnou, X.~N. Cao,
  E.~Dupoux, and A.-C. Bachoud-L{\'e}vi, ``Vocal markers from sustained
  phonation in huntington's disease,'' \emph{arXiv preprint arXiv:2006.05365},
  2020.

\end{thebibliography}


\end{document}